\documentclass[apj]{emulateapj}
\usepackage{times}
\usepackage{graphics,epsf}
\usepackage{amsmath}                
\usepackage{amsfonts}               
\usepackage{amssymb}                
\usepackage{epsfig}                 
\usepackage{graphicx}
\usepackage{url}
\usepackage{tabularx}
\usepackage{booktabs}
\usepackage{hyperref}
\usepackage[dvipsnames]{xcolor}

\def \cm{~\rm{cm}}
\def \s{~\rm{s}}
\def \ms{~\rm{ms}}

\def \km{~\rm{km}}

\def \g{~\rm{g}}

\def \erg{~\rm{erg}}
\def \foe{~\rm{foe}}

\def \foe{~\rm{foe}}

\begin{document}

\title{Jittering-jets explosion triggered by the standing accretion shock instability}

\author{Oded Papish}
\author{Avishai Gilkis}
\author{Noam Soker}

\affil{Department of Physics, Technion -- Israel Institute of
Technology, Haifa 32000, Israel;
 papish@physics.technion.ac.il; agilkis@tx.technion.ac.il; soker@physics.technion.ac.il}

\begin{abstract}
We show that the standing accretion shock instability (SASI) that
has been used to ease the shock revival in core collapse
supernovae (CCSNe) neutrino-driven explosion models, might play a much more decisive role in
supplying the stochastic angular momentum required to trigger an
explosion with jittering jets.
We find that if the kinetic energy associated with the transverse (non radial) motion of the SASI is larger than about ten percent of the energy associated with the energy of the accreted gas, then the 
stochastic angular momentum can reach about five percent of the Keplerian specific
angular momentum around the newly born neutron star.
Such an accretion flow leaves an open conical region along the
poles with an average opening angle of about 5 degrees. The outflow
from the open polar regions powers an explosion according to the
jittering-jets model.
\end{abstract}

\section{INTRODUCTION}
\label{sec:introduction}

Core-collapse supernovae (CCSNe) are explosions of massive stars.
A huge amount of gravitational energy, more than $10^{53} \erg$,
is released by the newly formed neutron star (NS), or black hole
(BH). The manner by which a small fraction of this energy is
channeled to explode the star is an open question. The two
contesting processes for exploding {\it all} CCSNe are the delayed
neutrino mechanism (e.g.,  \citealt{Wilson1985, bethe1985} and
\citealt{Janka2012} for a review) and the jittering-jets mechanism
\citep{Papish2011, Papish2012, PapishSoker2014a,
PapishSoker2014b, GilkisSoker2014}. Explosions based on jets
formed in cases with pre-collapse rapidly rotating cores exist as
well (e.g. \citealt{LeBlanc1970,Khokhlov1999,Lazzati2012}), but
these models can account for a limited number of rare types of CCSNe.

Another recent revisited model is the collapse-induced thermonuclear explosion (CITE) model \citep{Burbidgeetal1957,KushnirKatz2014}. In this model a helium-oxygen shell that is compressed during the collapse is detonated and then unbinds the outer stellar layers. The CITE model can result in up to a few$\times 10^{51} \erg$ of kinetic energy under very tuned parameters \cite{kushnir2015}.

The delayed-neutrino mechanism faces two challenges. The first one
is to revive the stalled shock of the inflowing core gas, and the
second challenge is to achieve the desired $\ga 10^{51} \erg = 1
\foe$ observed explosion kinetic energy. We note that highly energetic explosions (e.g., ASASSN-15lh,  \citealt{Dong2015}) cannot be explained by the neutrino mechanism while it can be accounted for with the jittering-jets model that is based on a negative feedback mechanism  \citep{Gilkisetal2016}. 
The incapability of the delayed-neutrino mechanism to overcome these two obstacles in a consistent and persistent manner is mirrored in the varying, and sometimes conflicting, outcomes of increasingly sophisticated
multidimensional core collapse simulations (e.g.,
\citealt{bethe1985,Burrows1985,Burrows1995,Fryer2002,Buras2003,
Ott2008,Marek2009,Nordhaus2010,Brandt2011,Hanke2012,Kuroda2012,
Hanke2012,Mueller2012,Bruenn2013,MuellerJanka2014,Mezzacappaetal2014,Mezzacappaetal2015}). For some other difficulties of the delayed neutrino mechanism see \cite{kushnir2015b}.

To ease the revival of the stalled shock in neutrino-based explosion  models,  dynamical effects, like pre-collapse convection and/or rotation, have
been studied in great details in recent years.
\cite{CouchOtt2013}, \cite{CouchOtt2015}, and
\cite{MuellerJanka2015} introduced pre-explosion turbulence in
the core. They found that after collapse the turbulence is carried
to the post-shock region, and an effective turbulent ram pressure
exerted on the stalled shock allows shock revival with less
neutrino heating. \cite{Abdikamalovetal2014},
however, find that increasing the numerical resolution allows a
cascade of turbulent energy to smaller scales, and the shock
revival becomes harder to achieve.

The main challenge of the jittering-jets model, on the other hand, is  to supply a large enough specific angular momentum to the mass accreted onto the NS to form an accretion disk or an accretion belt. A belt is
defined as a thick sub-Keplerian accretion disk that does not extend much
beyond the NS, but has sufficiently large specific angular
momentum to prevent an inflow along the two opposite polar
directions. \cite{GilkisSoker2015} showed that the above assumed
pre-collapse turbulence lead to the formation of intermittent
thick accretion disks, or accretion belts, around the newly born
NS. The implication of their results is that the pre-collapse
turbulence assumed by \cite{CouchOtt2013}, \cite{CouchOtt2015},
and \cite{MuellerJanka2015} facilitated much more the
jittering-jets model than the delayed neutrino mechanism.

Another dynamical effect that has been studied in relation to the
delayed neutrino mechanism is the standing accretion shock
instability (SASI) that develops in the post-shock inflowing core
material \citep{BlondinMezzacappa2003, BlondinMezzacappa2007,
Fernandez2010, Burrows1995, Janka1996, Buras2006a, Buras2006b,
Ott2008, Marek2009}. Most interesting to our present study
is the spiral modes of the SASI that includes transverse motion that
carries local angular momentum variations. The local variations
can add up to non-zero angular momentum. \cite{Rantsiouetal2011},
for example, suggested the spiral modes of the SASI as the source
of pulsar angular momentum. It was found that the spiral modes of
the SASI can reduce the neutrino flux that is required to revive
the stalled shock, e.g., \cite{Fernandez2015} and earlier
references therein.  We note that even if the stalled shock is
revived, the delayed neutrino mechanism encounters a severe
obstacle in achieving $1 \foe$ \citep{Papishetal2015}.

In the present paper we study the implications of the results of
\cite{Fernandez2015} on the jittering-jets model. As we show, the
SASI might play a significant role in facilitating the
jittering-jets model, hence might solve the biggest challenge 
of the jittering-jets model. In section \ref{sec:angular_momentum} we describe the way the opening angle along the two opposite polar directions is calculated. In section \ref{sec:belt} we calculate this angle from the results presented by \cite{Fernandez2010}. Our short summary is in section \ref{sec:summary}.

\section{ACCRETION BELT}
\label{sec:angular_momentum}

We consider a scenario where material falling on the proto-NS
has a temporary angular momentum in some direction, which we denote as the positive $z$-axis.
As the material possesses a specific angular momentum $j \ne 0$, 
 the accretion will be limited to some angle $\theta_a$ 
from the $z$-axis. 
This angle, $\theta_a$, can be estimated from the magnitude of the angular momentum 
by the balance between the centrifugal and gravitational forces. 
At a point on the NS surface and at an angle $\theta$ from the $z$-axis, 
the centrifugal force is $F_c=j^2/(R_{\rm NS} \sin \theta)^3 $
and the opposing gravitational component is
$F_G = {GM_\mathrm{NS} }\sin \theta /{R^2_\mathrm{NS}},$
where $M_\mathrm{NS}$ and $R_\mathrm{NS}$ are the proto-NS mass and radius, respectively.
The required specific angular momentum
for limiting the accretion to an angle $\theta_a$, is obtained by equating the two forces \footnote{
In \cite{GilkisSoker2015} there was erroneously a factor of $\sqrt{\sin\theta}$
in their Eq. (7), and the graphs presented results using a factor of $\sin\theta$
instead of $\sin^2\theta$,
which is equivalent to balancing entirely the gravitational force,
and not just the perpendicular component.
Using the correct expression results in higher specific angular momentum, and strengthens the conclusion of \cite{GilkisSoker2015}. }
\begin{equation}
j_z = \sqrt{GM_\mathrm{NS}R_\mathrm{NS}}\, \sin^2\theta_a
\label{eqreqj},
\end{equation}
and the limiting angle is
\begin{equation}
\theta_a = \sin^{-1} \sqrt{\frac{j_z}{j_\mathrm{Kep}}},
\label{eqangle}
\end{equation}
where $j_\mathrm{Kep} = \sqrt{GM_\mathrm{NS}R_\mathrm{NS}}$.

It is important to emphasize that the open polar regions (or  `avoidance regions' as they are avoidance regions for the incoming gas), do not serve to collimate the outflow. Even for thin accretion disks where the opening angle is close to $90^\circ$, e.g., as in young stellar objects, there are jets.  The role of the avoidance regions is to allow bipolar mass outflow as a result of the magnetic activity in the accretion belt (see below).  

Furthermore, as the open polar regions do not have a specific role, there is no threshold on their value. The collimated outflow is formed by magnetic activity where there are two opposite preferred directions (the rotation axis) along which the pressure of the inflowing gas is very low. The magnetic activity then leads to an outflow along these directions \citep{SchreierSoker2016}.  The situation is such that there is a monotonic relation between the low pressure of the incoming gas (which can be even zero) and the limiting angle. 
Numerical simulations are required to determine the value of the low pressure of the incoming gas that allows for an outflow to develop. Our estimate, that must be checked with 3D numerical simulations, is that for $\theta_a$ larger than about 0.1 (several degrees) a bipolar outflow  will develop.

This calculation of the limiting angle $\theta_a$ is under the assumption of a uniform specific accreted angular momentum $j=\left <j\right >$. In most cases the specific angular momentum is not uniform. Material with lower angular momentum can flow through the poles with an angle $\theta<\theta_a$, while material with higher angular momentum will form an  accretion belt with a higher limiting angle $\theta_a$ than what is assumed here. In general the limiting angle $\theta_a$ in equation \ref{eqangle} represents some average behavior. 

In \cite{GilkisSoker2015} this approach was used to show that if before collapse there exist high convective velocities in the progenitor, such as those presented by \cite{CouchOtt2013,CouchOtt2015} and  \cite{MuellerJanka2015}, those velocities can 
give rise to the required stochastic angular momentum needed for an accretion belt.

\begin{figure}[h!]
   \centering
    \includegraphics*[scale=0.55]{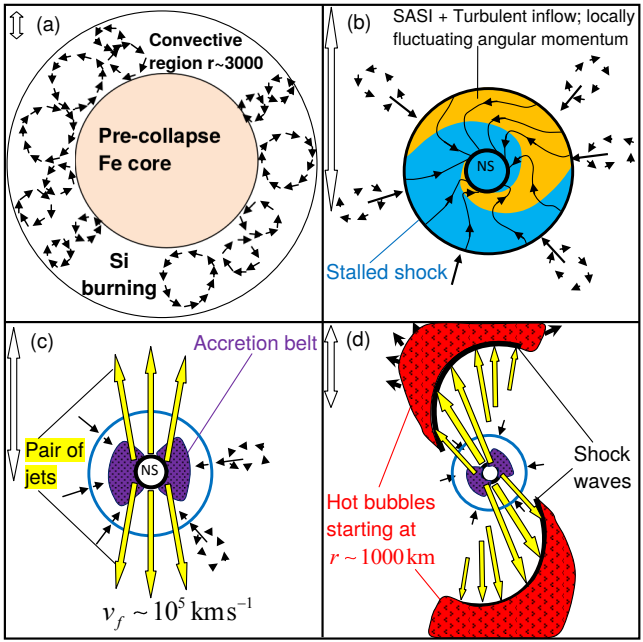} \\
\caption{ A schematic presentation of the proposed scenario.
The panels are not exactly to scale, but the two-sided
arrow on the upper left of each panel is approximately $500 \km$.
The four panels span an evolution time of several seconds. (a) In
the silicon burning shell of the pre-collapse core there is a
convective region, at about thousands of km from the center. The
convective vortices are a source of the stochastic angular
momentum. (b) After collapse and the formation of a neutron star
(NS) the rest of the in-falling gas passes through the stalled
shock. The stochastic spatial distribution of angular momentum in
the silicon burning shell is carried inward into the post-shock
region. In addition, the spiral modes of the SASI add stochastic angular momentum in the post-shock region.(c) The accreted angular momentum changes stochastically in magnitude and direction. For short periods of times, tens of
milliseconds, the accreted gas near the NS possesses a net angular
momentum. Accretion along and near the temporary poles of the
angular momentum axis is inhibited, and a temporary accretion belt
is formed around the newly born NS. If the belt exists for a long
enough time, several dynamical times, or $> 0.01 \s$, it can spread
in the radial direction to form an accretion disk. The belt or
disk are assumed to launch two opposite jets with initial
velocities of $v_f \approx 10^5 \km \s^{-1}$ (about the escape
velocity from the newly formed NS). (d) The jets that are launched
in varying directions, called jittering jets, penetrate through
the gas close to the center, and their shocked gas inflate hot
bubbles (see \citealt{Papish2011}). These bubbles expand and
explode the star in the jittering jets model
\citep{PapishSoker2014a, PapishSoker2014b}.}
      \label{fig:schem}
\end{figure}

In this work we consider the stochastic angular momentum resulting from post-bounce dynamical SASI instabilities.
The schematic flow structure discussed here, including stochastic angular momentum from the pre-collapse core and the SASI, is presented in Figure \ref{fig:schem}.

\section{ACCRETION BELTS FROM THE SASI SPIRAL MODES}
\label{sec:belt}
\cite{Fernandez2010} studied the spiral modes of the SASI using 3D simulations.
He found that the SASI leads to a redistribution of the angular momentum accreted onto the proto-NS. 
This angular momentum was less than the Keplerian angular momentum close to the proto-NS and no accretion disk was formed. 
In this section we revisit the results of \cite{Fernandez2010} 
and show that the SASI can lead to a belt like structure around the proto-NS. 
We speculate that as a result of the belt like flow, jets will be launched and explode the star \citep{SchreierSoker2016}. 

The angular momentum accreted onto the proto-NS in the simulations conducted by  \cite{Fernandez2010} can be estimated from the rate of change of the proto-NS rotational period $T$,
as presented in his figure 17 for five cases. 
The rate of change of the angular momentum near the proto-NS is 
\begin{equation}
\dot J = I \dot \omega = -I \frac{2\pi}{T^2} \dot T.
\end{equation}  
For the mass inflow rate of $ \dot M_{\rm acc} \approx 0.3 M_\odot \s^{-1}$ and a NS moment of inertia $I=10^{45} \g \cm^2$ used in the simulations, we can estimate the specific angular momentum of the accreted mass as function of time , 
\begin{equation}
 \textstyle
j=1.75 \times 10^{15} \left( \frac{\dot J}{7 \times 10^{45}\g \cm^2 \s^{-1}} \right ) \left (\frac{\dot M_{\rm acc}}{0.3 M_\odot \s^{-1}}\right )^{-1} \cm^2 \s^{-1},
\label{eq:j}
\end{equation}
using the results of \cite{Fernandez2010}.

\begin{figure}
   \centering
    \includegraphics[scale=0.5]{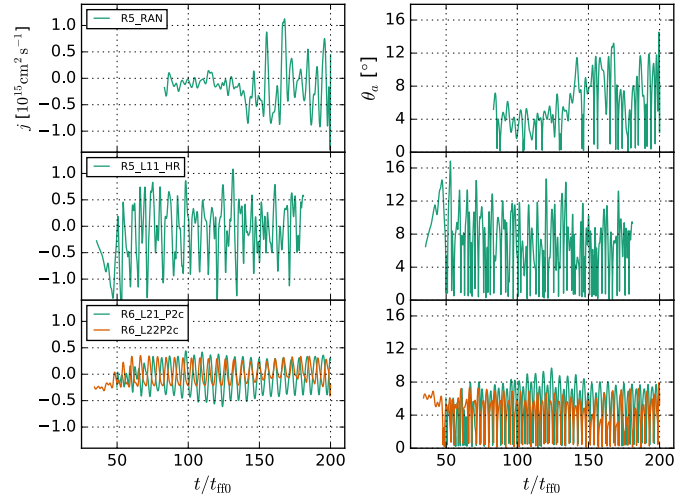} \\
\caption{Left panels: the specific angular momentum $j$ as calculated from equation (\ref{eq:j}) with the orbital period of the NS taken from figure 17 of \cite{Fernandez2010}. Right panels: the limiting angle $\theta_a$ according to equation (\ref{eqangle}), and for the specific angular momentum values from the left panel of each case. The four different models are digitized from \cite{Fernandez2010} with $t_{\rm ff0}\simeq3.1 \ms$., where details of the models can be found }
      \label{fig:SASI}
\end{figure}

The accreted specific angular momentum $j$ calculated by equation (\ref{eq:j}) is presented in the left panels of Figure. \ref{fig:SASI}. The values of $\dot M_{\rm acc}$ and $I$ are as scaled in equation (\ref{eq:j}) , while $T$ and $\dot T$ are from Figure 17 of \cite{Fernandez2010}. We also plot the limiting angle $\theta_a$ calculated from equation (\ref{eqangle}), with the scaling of 
\begin{equation}
j_{\rm Kep}=2.16 \times 10^{16} \left( \frac{M}{1.4 M_\odot} \right )^{1/2} \left (\frac{R}{25 \km}\right )^{1/2} \cm^2 \s^{-1}
\label{eq:kepler}
\end{equation}
for the same cases in the right panels of the same figure. As can be seen, the specific angular momentum is indeed lower by an order of magnitude than that required to form an accretion disk around the proto-NS. However, we find the limiting angle, $\theta_a$ to be large enough to create a belt like structure around the proto-NS in most cases. 

{
Let us dwell on some of the ingredients of the proposed mechanism. In a recent paper \cite{Mostaetal2015} conducted very high resolution simulations of CCSNe with pre-collpase rapidly rotating cores. They showed that rapidly rotating material around the newly born NS can substantially amplify magnetic fields, with an e-folding time scale of $\tau_e \approx 0.5 \ms$. In their simulations this is about half an orbital period in the relevant region of the disk.
}

It is important to note the following properties of the results obtained by  \cite{Mostaetal2015}.
(1) \cite{Mostaetal2015} obtained significant magnetic field amplification only for very high spatial
resolution simulations. 
 
(2) The amplification reaches saturation when the magnetic energy density is about equal to the turbulent energy density (equipartition). In their simulations this occurs within 3ms. Had the initial magnetic field been weaker, amplification would have last longer, still reaching equipartition. 
(3) The amplification time at a radius of about $40 \km$ is about ten times as long as in the inner radius. This increase in amplification time results from two factors. Firstly, the Keplerian orbital period at $40 \km$ is larger by a factor of about 4 relative to that at a radius of  $15 \km$.  Secondly, the shear is large near the NS.  
In the scenario proposed by \cite{SchreierSoker2016} the amplification occurs near the surface of the NS, hence the amplification time of the magnetic field is expected to be short. 

The Keplerian orbital period at $\sim 25 \km $ from the newly born NS is $1.8 \ms$. From Fig. \ref{fig:SASI} we see that a typical temporary disk last for about $5-10 \ms$, that is 3-6 times the orbital period. From the results of  \cite{Mostaetal2015} the magnetic fields can be amplified by $\approx  \exp(5-12) =100-10^5$.

According to \cite{SchreierSoker2016} it is the amplification of the magnetic field that is the most important ingredient in the launching of jets from accretion belts. The second parameter in importance is the opening angle $\theta_a$. The reason for the higher importance of the magnetic fields is that the magnetic activity can change the opening angle in the following ways.

\cite{SchreierSoker2016} suggest that  reconnection of the magnetic field lines eject gas through the two opposite polar avoidance regions. This activity can increase the opening angle in the inflowing gas. \cite{SchreierSoker2016} further argue that winding of the magnetic field lines frozen to the polar outflow  can further channel rotation energy to outflow kinetic energy. Magnetic tension can further increase the opening angle.  The main conclusion is that once the magnetic field become strong, the opening angle is opened to $\theta_a \ga 0.1 = 6^\circ$.

There is no upper limit on the value of $\theta_a$, as the scenario for jet launching from accretion belts does not require the belt to collimate the bipolar outflow.

\section{SUMMARY}
\label{sec:summary}

In recent years dynamical effects, like pre-collapse convection and/or rotation, have been introduced in simulations of neutrino-driven explosion of CCSNe. The hope was that these effects will help revive the stalled shock, and lead to the desired explosion energy that has not been consistently achieved with neutrino-based mechanisms. 
The dynamical effects include the SASI (e.g., \citealt{ BlondinMezzacappa2003, BlondinMezzacappa2007,Fernandez2010}) and pre core-collapse perturbations and turbulence (e.g.,\citealt{CouchOtt2013, CouchOtt2015,MuellerJanka2015}).
These dynamical effects have been shown to have onlyklimited effects on helping a successful explosion by the delayed-neutrino mechanism. However, 
these effects might help in creating accretions belts and so might be important ingredients in the jittering-jets model for CCSNe.

In a former paper \cite{GilkisSoker2015} studied the influence of pre-collapse core turbulence 
on the jittering-jets model, and found it to help with supplying the stochastic angular momentum. In this paper we study the influence of the SASI on the creation of intermittent accretion belts. The schematic flow structure is presented in Figure \ref{fig:schem}. To calculate the specific angular momentum of the accreted mass  we use the study of the SASI spiral modes conducted by \cite{Fernandez2010}. 

We found that during many time intervals the average specific angular momentum of the accreted mass is $\approx 5 \%$ of the Keplerian angular momentum on the equator of the newly formed NS (left panels of Figure \ref{fig:SASI}). This implies that a cone with an angle of $\theta_a \approx 10^\circ$ from the temporary angular momentum axis will be almost devoid of accreted gas close to the NS. The temporal variations of the angle $\theta_a$ according to four cases of the SASI studied by  \cite{Fernandez2010}, are given in the right panels of Figure \ref{fig:SASI}. If magnetic fields are amplified in the accretion belt, due to sheared rotation and converging accretion flow, jets might be launched along the empty polar cones \citep{SchreierSoker2016}. This is a basic assumption of the jittering-jets model.

We note that the CITE thermonuclear explosion mechanism for CCSNe studied
by \cite{KushnirKatz2014} and \cite{kushnir2015} requires a large amount of angular momentum in the core to achieve the desired explosion energy from the thermonuclear burning of the mixed helium-oxygen layer. 
The collapsing rapidly rotating core supplies a vast amount of mass to form an accretion disk around the newly formed NS; about $1
M_\odot$ with a specific angular momentum of $j \approx 4 \times
10^{17} \cm^2 \s^{-1}$. The energy carried by the expected jets
will dwarf the energy released by the thermonuclear burning  \citep{Gilkisetal2016}.

Although the results of this paper and \cite{GilkisSoker2015} are only preliminary, they show that it might be possible to achieve the conditions for jets launching in CCSNe. If this is correct then the jittering-jets model will be able to explode a star with the desiered $1 \foe$ explosion energy\citep{PapishSoker2014b}. There are many more points that should be addressed before we can claim more conclusively that the jittering-jets model can work. This include simulations of magnetic fields amplification in the accretion belt and farther investigating the ability of it to launch jets.

This research was supported by a generous grant from the president of the Technion Prof. Peretz Lavie.


\label{lastpage}

\end{document}